\documentstyle[11pt,psfig,twoside]{article}

\begin{document}

\newcommand{\dd}{\,{\rm d}}
\newcommand{\ie}{{\it i.e.},\,}
\newcommand{\etal}{{\it et al.\ }}
\newcommand{\eg}{{\it e.g.},\,}
\newcommand{\cf}{{\it cf.\ }}
\newcommand{\vs}{{\it vs.\ }}
\newcommand{\zdot}{\makebox[0pt][l]{.}}
\newcommand{\up}[1]{\ifmmode^{\rm #1}\else$^{\rm #1}$\fi}
\newcommand{\dn}[1]{\ifmmode_{\rm #1}\else$_{\rm #1}$\fi}
\newcommand{\upd}{\up{d}}
\newcommand{\uph}{\up{h}}
\newcommand{\upm}{\up{m}}
\newcommand{\ups}{\up{s}}
\newcommand{\arcd}{\ifmmode^{\circ}\else$^{\circ}$\fi}
\newcommand{\arcm}{\ifmmode{'}\else$'$\fi}
\newcommand{\arcs}{\ifmmode{''}\else$''$\fi}
\newcommand{\MS}{{\rm M}\ifmmode_{\odot}\else$_{\odot}$\fi}
\newcommand{\RS}{{\rm R}\ifmmode_{\odot}\else$_{\odot}$\fi}
\newcommand{\LS}{{\rm L}\ifmmode_{\odot}\else$_{\odot}$\fi}

\newcommand{\Abstract}[2]{{\footnotesize\begin{center}ABSTRACT\end{center}
\vspace{1mm}\par#1\par
\noindent
{~}{\it #2}}}

\newcommand{\TabCap}[2]{\begin{center}\parbox[t]{#1}{\begin{center}
  \small {\spaceskip 2pt plus 1pt minus 1pt T a b l e}
  \refstepcounter{table}\thetable \\[2mm]
  \footnotesize #2 \end{center}}\end{center}}

\newcommand{\TableSep}[2]{\begin{table}[p]\vspace{#1}
\TabCap{#2}\end{table}}

\newcommand{\FigCap}[1]{\footnotesize\par\noindent Fig.\  %
  \refstepcounter{figure}\thefigure. #1\par}

\newcommand{\TableFont}{\footnotesize}
\newcommand{\TableFontIt}{\ttit}
\newcommand{\SetTableFont}[1]{\renewcommand{\TableFont}{#1}}

\newcommand{\MakeTable}[4]{\begin{table}[htb]\TabCap{#2}{#3}
  \begin{center} \TableFont \begin{tabular}{#1} #4 
  \end{tabular}\end{center}\end{table}}

\newcommand{\MakeTableSep}[4]{\begin{table}[p]\TabCap{#2}{#3}
  \begin{center} \TableFont \begin{tabular}{#1} #4 
  \end{tabular}\end{center}\end{table}}

\newenvironment{references}%
{
\footnotesize \frenchspacing
\renewcommand{\thesection}{}
\renewcommand{\in}{{\rm in }}
\renewcommand{\AA}{Astron.\ Astrophys.}
\newcommand{\AAS}{Astron.~Astrophys.~Suppl.~Ser.}
\newcommand{\ApJ}{Astrophys.\ J.}
\newcommand{\ApJS}{Astrophys.\ J.~Suppl.~Ser.}
\newcommand{\ApJL}{Astrophys.\ J.~Letters}
\newcommand{\AJ}{Astron.\ J.}
\newcommand{\IBVS}{IBVS}
\newcommand{\PASP}{P.A.S.P.}
\newcommand{\Acta}{Acta Astron.}
\newcommand{\MNRAS}{MNRAS}
\renewcommand{\and}{{\rm and }}
\section{{\rm REFERENCES}}
\sloppy \hyphenpenalty10000
\begin{list}{}{\leftmargin1cm\listparindent-1cm
\itemindent\listparindent\parsep0pt\itemsep0pt}}%
{\end{list}\vspace{2mm}}

\def\TYLDA{~}
\newlength{\DW}
\settowidth{\DW}{0}
\newcommand{\dw}{\hspace{\DW}}

\newcommand{\refitem}[5]{\item[]{#1} #2%
\def\REFARG{#3}\ifx\REFARG\TYLDA\else, {\it#3}\fi
\def\REFARG{#4}\ifx\REFARG\TYLDA\else, {\bf#4}\fi
\def\REFARG{#5}\ifx\REFARG\TYLDA\else, {#5}\fi.}

\newcommand{\Section}[1]{\section{#1}}
\newcommand{\Subsection}[1]{\subsection{#1}}
\newcommand{\Acknow}[1]{\par\vspace{5mm}{\bf Acknowledgements.} #1}
\pagestyle{myheadings}

\begin{center}

{\large\bf Early-Type Contact Systems in the LMC MACHO Database}
\vskip0.8cm
{\bf S.~M.~~ R~u~c~i~n~s~k~i}
\vskip3mm
{David Dunlap Observatory, University of Toronto
P.O.~Box~360, Richmond Hill, Ontario L4C~4Y6, Canada\\
e-mail: rucinski@astro.utoronto.ca}

\end{center}

\Abstract{Two sub-samples of blue, luminous contact systems from among 86 
systems discovered in LMC by the MACHO project and classified as EB3 have been 
analyzed: a sub-sample of 29 extremely-blue systems (XB) with the observed 
colors ${V-R_C<-0.1}$ and a sub-sample of 36 moderately-blue (MB) systems with 
${-0.1<V-R_C<0}$. To be so blue, the XB systems must be intrinsically very hot 
and almost unreddened so that lack of information on reddening is unimportant 
for them; their properties offer us a first-time insight into the 
absolute-magnitude calibration for massive contact binaries. It has been found 
that the LMC systems are apparently not limited by the blue, short-period 
envelope in the color-period diagram which had been established on the basis 
of systems observed in the solar neighborhood, in galactic open clusters and 
in the sample of Old Disk systems in the direction of the Galactic Bulge. The 
period--magnitude correlation for the XB systems has a similar slope to those 
established in the absolute-magnitude, ${B-V}$ and ${V-I_C}$ based, 
calibrations for W~UMa-type systems.}{~} 

\Section{Introduction}
Calibrations of the form ${a_P\log P+a_C{\rm color}+a_0=M_V}$ are relatively 
successful in predicting absolute magnitudes of contact binaries of the 
W~UMa-type. They have been established on the basis of systems in open 
clusters (Rucinski 1994a) and then improved for the solar neighborhood using 
the Hipparcos data (Rucinski and Duerbeck 1997); with metallicity corrections, 
they can be used for globular clusters as well (Rucinski 1994b, 1995). They 
have been invaluable in handling the Galactic Disk and open cluster data for 
contact binaries (Rucinski 1998a). It would be very useful to establish a 
similar calibration for contact binaries of early spectral types because such 
binaries are now being detected in large numbers in nearby galaxies 
(Ka{\l}u\.{z}ny \etal 1998, Stanek \etal 1998, Stanek \etal 1999) and may 
eventually provide independent distance estimates. 

Contact binaries of spectral earlier than middle-A and periods longer than 
about 1.5 day are very rare in the Galaxy. They practically do not exist in 
the OGLE Galactic-Bulge database which probes mostly the Old Disk population 
(Rucinski 1997a, 1998a); the period distribution of contact binaries abruptly 
ends at about 1.3--1.5 days (Rucinski 1998b). This cut-off can be explained 
by non-existence of Old Disk main-sequence stars with masses much larger than 
the solar mass. Intrinsically bright, massive contact binaries, such as the 
systems discussed by Popper (1982) -- with periods of a few days and masses of 
several solar masses -- do exist within our Galaxy, but their frequency of 
occurrence is apparently exceedingly low and currently unknown. It is 
important that they share the same property of identical effective 
temperatures of components -- in spite of strongly differing masses -- with 
the solar-type contact binaries of the W~UMa-type. This property basically 
defines the W~UMa-type stars. 

The currently ongoing microlensing projects lead to discoveries of large 
numbers eclipsing binaries in Magellanic Clouds. This paper presents a pilot 
study based on the data for contact binaries in the Large Magellanic Cloud 
(LMC) which have been collected by the MACHO project (Alcock \etal 1997). In 
the next sections, we describe various properties of the early-type contact 
systems in the MACHO sample. The electronically-available database (American 
Astronomical Society CD-ROM Series, Vol.~8, 1997) contains the summary, 
time-independent data and plots of the light curves, and is quite typical for 
microlensing projects. It can be treated similarly as the one obtained by the 
OGLE project for the direction toward the Galactic Bulge (Rucinski 1997a). In 
this spirit, only the mean quantities such as the period, the maximum and the 
in-eclipse $V$-magnitudes as well as the maximum-light ${V-R_C}$ color index 
in the Kron--Cousins system have been used. Availability of this particular 
index is somewhat unfortunate because the previous absolute-magnitude 
calibrations for W~UMa-type systems were based on the ${B-V}$ and ${V-I_C}$ 
indices so that the existing calibrations can be used only for general 
guidance, but not for detailed comparisons. 

\vspace*{9pt}
\Section{The Sample} 
\vspace*{4pt}
The definition of W~UMa-type contact systems (normally abbreviated as EW) does 
not include binaries with periods longer than one day. In the MACHO database 
of the eclipsing binaries discovered in LMC (Alcock \etal 1997) such systems 
are called EB, with a subset of them, with eclipses of similar depth 
indicating similar surface brightness of components, designated as EB3. The  
EB3 systems have been used in this paper, keeping in mind subjectivity of this 
definition. Electronically available plots of the light curves confirm, that 
the light-curve shapes are indeed very typical for contact binaries, as shown 
for the example in Fig.~7 in Alcock \etal (1997). There are 86 systems of this 
type in the MACHO database. 

In addition to the values of the orbital periods, the photometric data are 
important for discussions of this paper. The precision of the MACHO data is 
moderate: The mean standard error of $V$ magnitudes is about 0.07 while the 
mean standard error for the color index ${V-R_C}$ is about 0.03. The accuracy 
(which reflects the role of systematic errors) is limited for the MACHO sample 
by the use of the non-standard bandpasses which are not defined -- as usually 
-- by glass filters, but by a dichroic filter separating the bands, and by 
transmission characteristics of the CCD's and of the atmosphere. 

\begin{figure}[htb]
\centerline{\psfig{figure=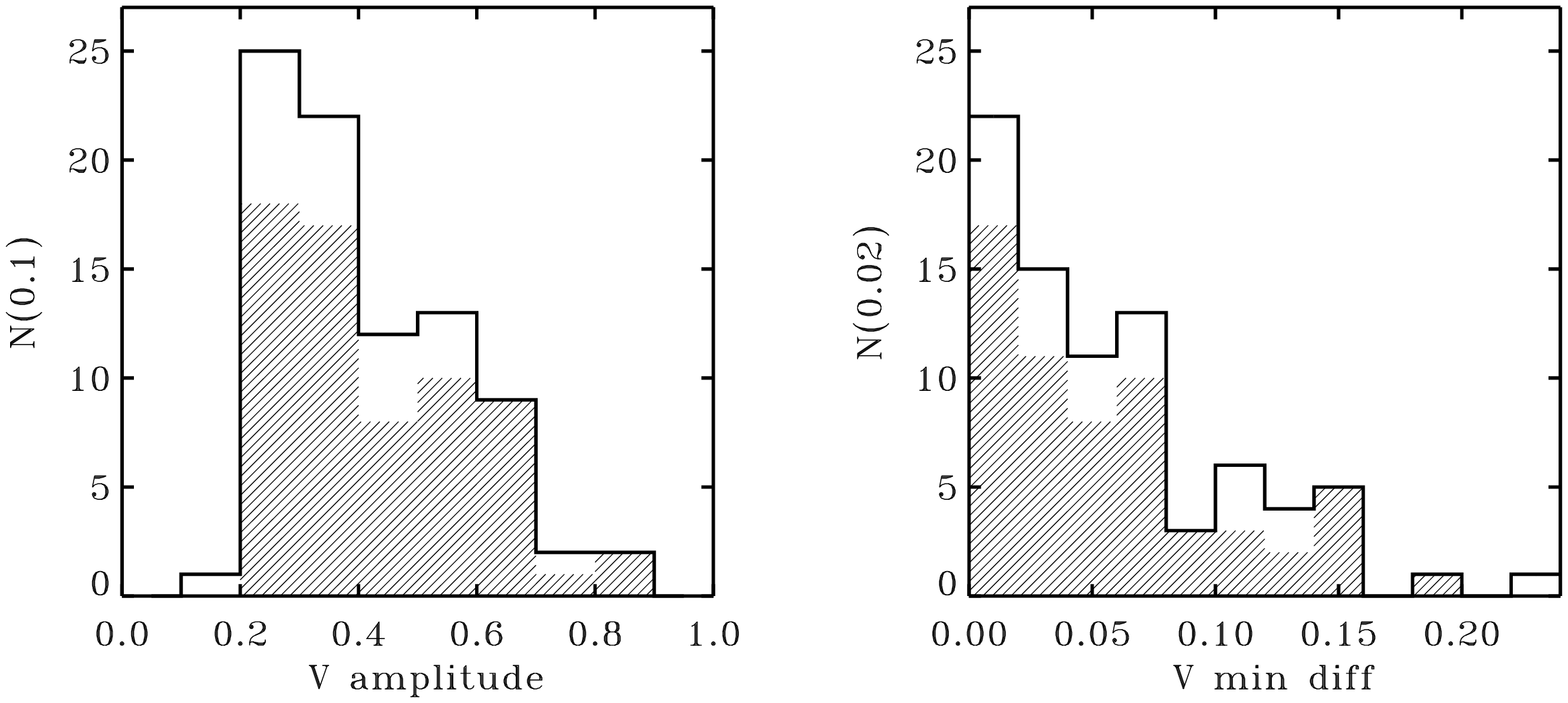,bbllx=110pt,bblly=80pt,bburx=450pt,bbury=770pt,width=13cm,angle=90,clip=}}
\FigCap{Histograms of the amplitudes (left panel) and eclipse-depth 
differences (right panel) in $V$ magnitudes for the whole sample of 86 EB3 
systems (continuous line) and for the sub-sample of blue systems with the 
observed color index ${V-R_C<0}$ (shaded area).} 
\end{figure}
The histograms of the $V$-magnitude amplitudes and minimum differences shown 
in Fig.~1 are quite typical for contact systems. The largest amplitudes 
observed are slightly over 0.8~mag and the minimum differences are typically 
in the range of 0 to 0.02~mag, as for genuine W~UMa-type systems (see Fig.~2 
in Rucinski 1997b; the magnitude difference is approximately twice the Fourier 
coefficient $a_1$ shown there). The limit for inclusion in the database was 
the variability amplitude larger than 0.2~mag. Contact binaries with still 
smaller variability amplitudes are expected to be actually the most frequent 
for randomly distributed orbital inclinations (Rucinski 1997b). 

\Section{The Color--Magnitude Diagram (CMD)}
The CMD for the contact systems considered here shows a concentration of the 
\begin{figure}[htb]
\vspace*{-14pt}
\centerline{\psfig{figure=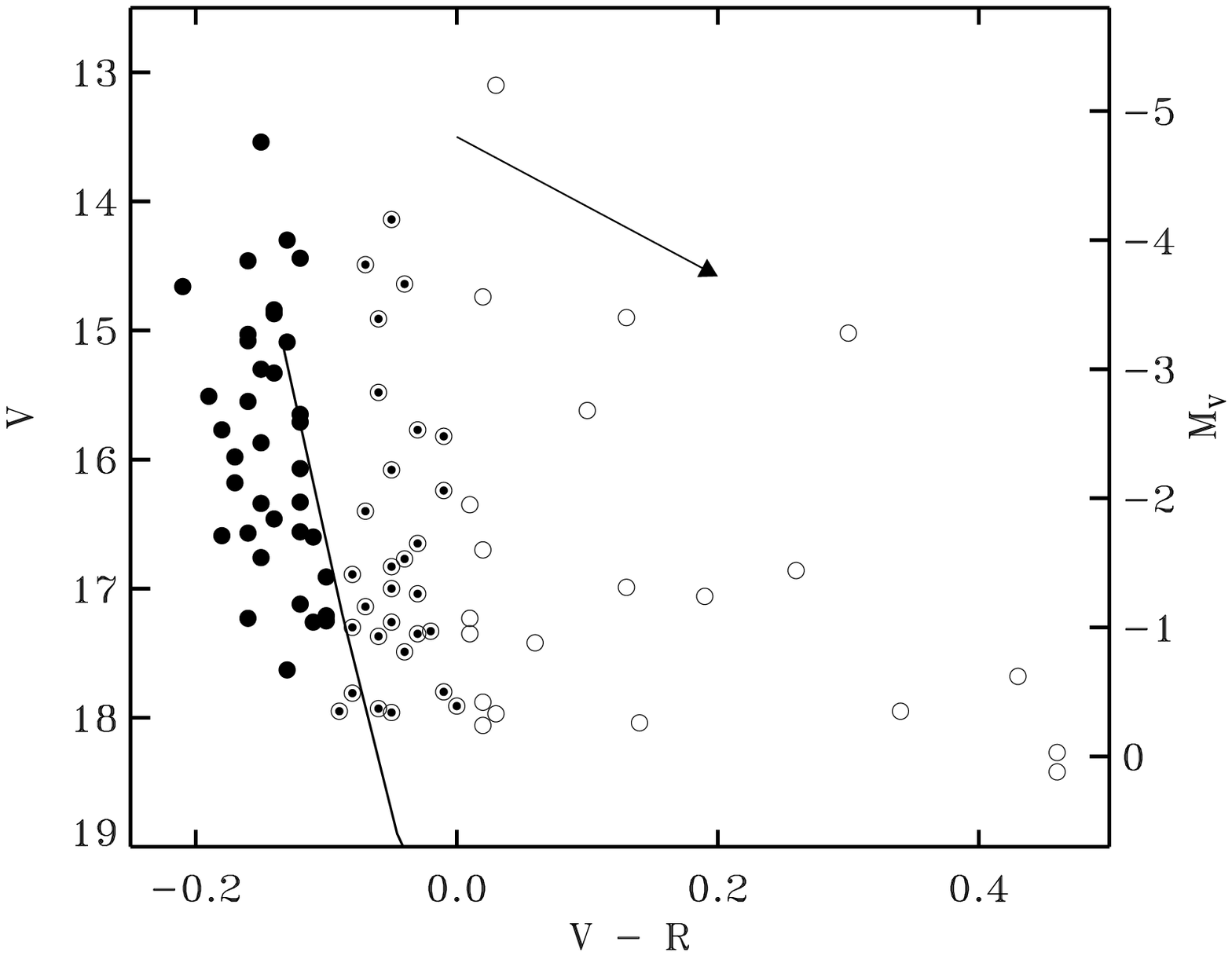,bbllx=55pt,bblly=320pt,bburx=555pt,bbury=720pt,width=11cm,clip=}}
\FigCap{The color--magnitude diagram for the MACHO sample of contact binaries. 
The limits at ${V-R_C=-0.1}$ and 0 define three ranges that we consider in the 
paper. The binaries bluer than ${V-R_C=-0.1}$ (XB group) are marked by filled 
circles while those redder than ${V-R_C=0}$ are marked by open circles; the 
semi-filled circles are for ${-0.1<V-R_C<0}$ (MB group). The bluest objects 
are those that are intrinsically hottest and least reddened. The curve at the 
left edge gives the Zero Age Main Sequence following Schmidt-Kaler (1982); it 
is shown without any correction for interstellar reddening. For simplicity, 
the distance modulus ${m-M=18.3}$ was assumed, but this value is not used in 
the paper for any purpose. The arrow gives the reddening vector ${A_V=5.4 
E_{V-R_C}}$.} 
\end{figure}
systems at the left, blue edge of the diagram and a scattered population of 
systems in the red part of the diagram (Fig.~2). Since the available data are 
in two photometric bands only, we have no information about reddening of 
individual objects. Reddening in LMC is patchy and varies within ${0<E_{B-V} 
<0.4}$, but with excursions up to 0.8 (Harris \etal 1997). In this situation, 
it has been decided to limit our considerations to the systems at the left 
edge of the CMD which are intrinsically the bluest and least-reddened. The 
systems in the red part of the diagram in their majority have long orbital 
periods (see the next Section and Fig.~3) and we do not have a good 
explanation for them. In what follows, we will consider three groups of the 
contact binaries defined by the observational ${V-R_C}$ color ranges: the 
extremely blue (XB) systems with ${V-R_C<-0.1}$, the moderately blue (MB) 
systems with ${0.1<V-R_C<0}$ and the red systems with ${V-R_C>0}$; the latter 
are not discussed here at all. 

\begin{figure}[htb]
\vspace*{-.4cm}
\centerline{\psfig{figure=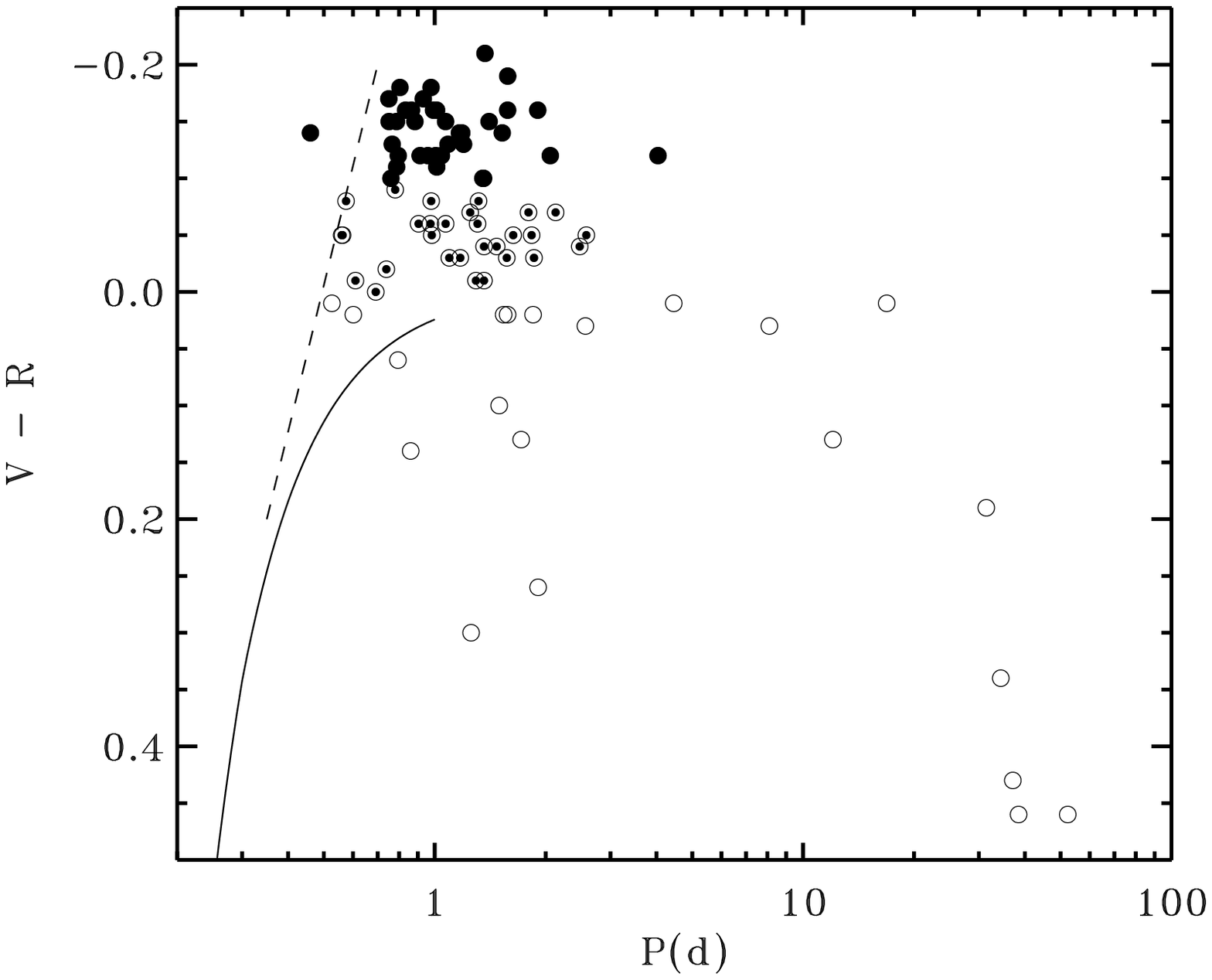,bbllx=55pt,bblly=320pt,bburx=555pt,bbury=720pt,width=11cm,clip=}}
\vspace*{-.2cm}
\FigCap{The period-color diagram with symbols as in the previous figure. The 
curved line is the Short-Period Blue Envelope (SPBE) previously found to limit 
location of least-evolved contact binaries in the Galaxy. The broken line 
gives a suggested location of the SPBE on the assumption that the  Galactic 
sample did not include sufficiently un-evolved objects.} 
\end{figure}

\Section{The Period--Color Diagram (PCD)}
The period-color diagram is shown in Fig.~3, with the same color-range symbols 
as in Fig.~2. The curved line at the left edge gives the Short-Period 
Blue-Envelope (SPBE) for the Galactic-Disk sample, ${(V-I_C)=0.053P^{-2.1}}$ 
(Rucinski 1997a), after application of the color-color transformation to 
${V-R_C}$ using Table~3 in Taylor (1986). It is obvious that the LMC systems 
are seen beyond the blue limit of the SPBE which was previously found to obey 
for all contact binaries in the galactic field and in all open clusters. The 
low metallicity of LMC (${\rm [Fe/H]=-0.5}$) cannot produce the shift because 
the intrinsic color indices for main-sequence OB stars have a negligible 
dependence on metallicity (Oestreicher \etal 1995). We strongly suspect that 
by considering the bluest contact systems, we are also looking at objects of a 
young stellar population which simply does not exist in the solar neighborhood 
or in old open galactic clusters. It is possible, that the curvature of the 
SPBE that was found before was entirely due to the lack of young objects in 
the previously-used samples; possibly, the real location of the SPBE is closer 
to what has been marked by a broken line in Fig.~3. 

A closer inspection of the PCD in Fig.~3 shows that the blue contact systems 
in LMC have orbital periods up to 2--3 days, whereas in the disk sample of 
OGLE the period distribution was found to a sharp cut-off at about 1.3--1.5 
days (Rucinski 1998b). There are too few systems in the MACHO sample to 
analyze statistically the numbers of systems with periods above 1.3 days, 
however. 

\Section{The Period--Luminosity Relation}
The contact systems in LMC are practically in the same distance from us so 
that the initial goal of this study was an attempt to establish an 
absolute-magnitude calibration of the type ${a_P\log P+a_{VR}(V-R_C)+a_0 
=M_V}$. However, this goal is impossible to achieve at this time for the 
following reasons: (1)~Lack of reddening information prevents determination of 
reddening and absorption corrections for individual systems; (2)~${V-R_C}$ 
color index, as other ones utilizing optical spectral bands, loses sensitivity 
to the effective temperature for very hot stars and hence must be determined 
with high accuracy; (3)~The observed color indices have modest precision of 
about 0.03~mag and an uncharacterized accuracy due to the use of the 
non-standard bandpasses. At present, one can only address the matter of period 
dependence in the observed values of $V_{\rm max}$. Such a limited goal is 
still a valuable check on the assumptions because a period dependence is 
expected only for genuinely contact systems and it is by no means obvious that 
the systems classified as EB3 are indeed contact ones. For detached binaries, 
no relation between period and component brightness is expected. 

\begin{figure}[htb]
\centerline{\psfig{figure=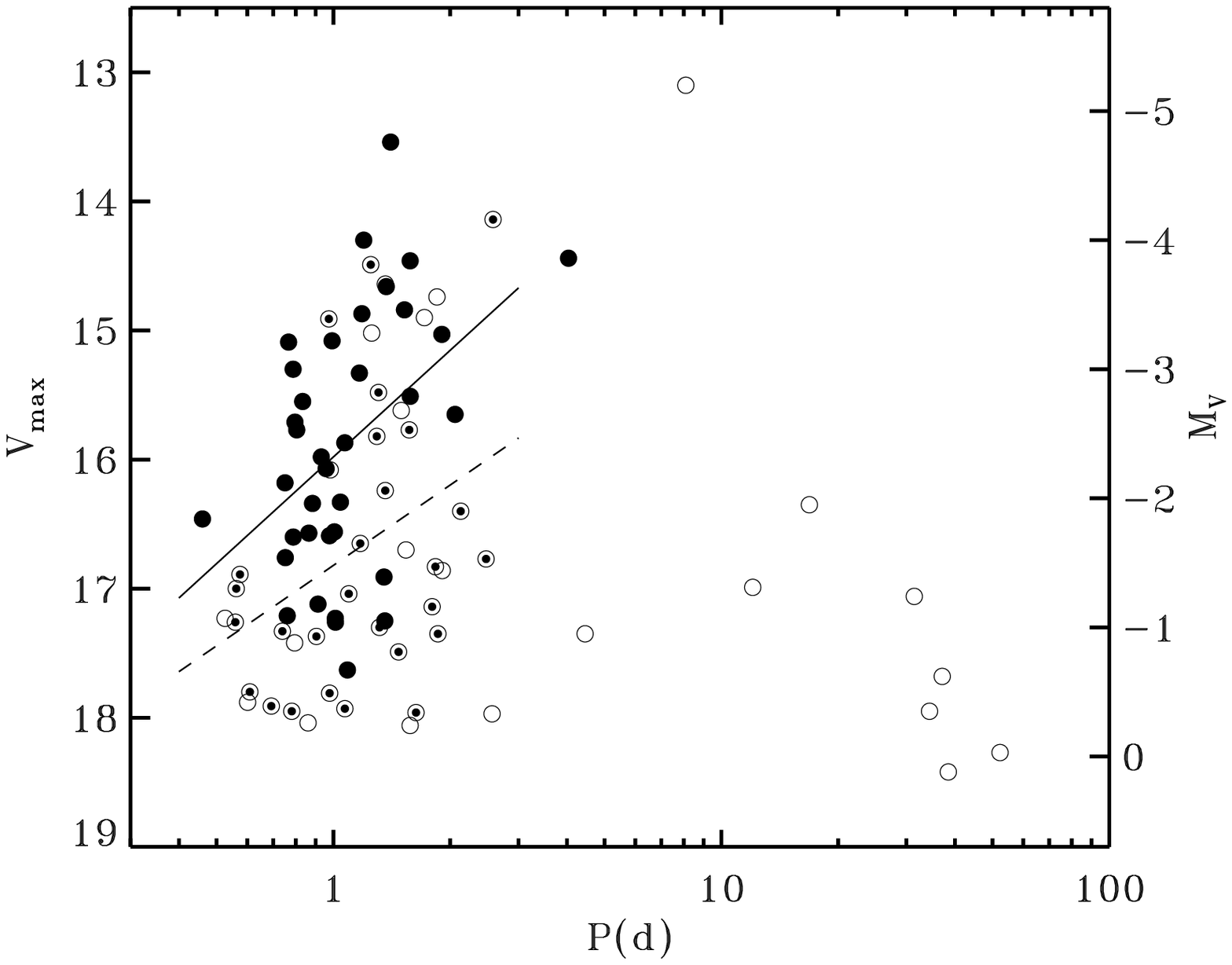,bbllx=55pt,bblly=320pt,bburx=555pt,bbury=720pt,width=11cm,clip=}}
\vspace*{-.2cm}
\FigCap{The relation between the orbital period and the $V$-magnitude at 
maximum light for the systems of the MACHO sample. The same symbols as in the 
previous figures are used to distinguish the three color index ranges. The 
lines give linear fits to the XB (solid line) and MB (broken line) 
sub-samples.} 
\end{figure}
The period -- observed-magnitude diagram for the sample is shown in Fig.~4. As 
one can see, the systems of both groups, the extra-blue (XB) and the 
moderately-blue (MB) systems, show some period dependence, but the scatter of 
points is large, probably mostly because of the unaccounted effects of the 
interstellar reddening and extinction. We note also that there are more MB 
systems in the vicinity of the faint-limit of the sample at ${V_{\rm max} 
\simeq18}$ than XB systems. Least-squares fits of the form ${V_{\rm max}=a_0 
+a_1\log P}$ have been performed for both samples; we note that the orbital 
periods $P$ are known practically without errors. Because the scatter of the 
data is large, uncertainties of the coefficients crucially depend on the data 
sampling. To characterize this effect, the errors of the coefficients $a_i$ 
have been determined using the bootstrap re-sampling technique. The results 
are listed in Table~1 in terms of the median values and the 68 percent (for 
Gaussian distributions, ${\pm1}$-sigma) and 95 percent (${\pm2}$-sigma) 
confidence levels. 
\MakeTable{ccccc}{12.5cm}{Bootstrap results and significance ranges
for coefficients of the fits: $V_{max} = a_0 + a_1 \log P$}
{
\hline
& \multicolumn{2}{c}{XB} & \multicolumn{2}{c}{MB} \\
& $a_0$ & $a_1$          & $a_0$ & $a_1$\\ 
\hline
$-95$\% ($-2\,\sigma$) & 15.69 & $-4.47$ & 16.45 & $-4.09$ \\
$-68$\% ($-1\,\sigma$) & 15.83 & $-3.43$ & 16.65 & $-3.07$ \\
median                 & 15.98 & $-2.74$ & 16.82 & $-2.07$ \\
$+68$\% ($+1\,\sigma$) & 16.12 & $-2.17$ & 17.00 & $-1.24$ \\
$+95$\% ($+2\,\sigma$) & 16.28 & $-1.36$ & 17.16 & $-0.37$ \\
\hline
}

\begin{figure}[htb]
\centerline{\psfig{figure=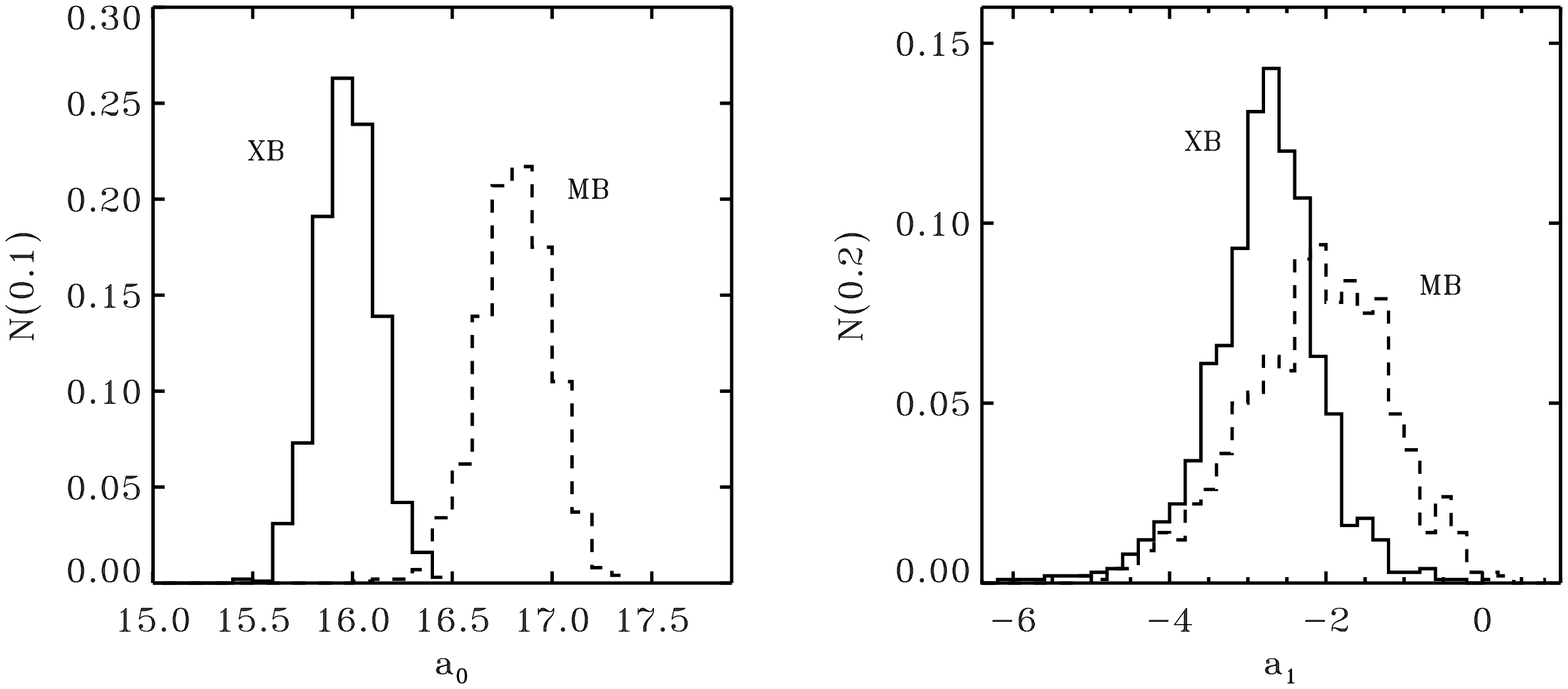,bbllx=110pt,bblly=80pt,bburx=450pt,bbury=770pt,width=13cm,angle=90,clip=}}
\FigCap{Histograms of the coefficients $a_0$ and $a_1$ in the fits ${V_{\rm 
max}=a_0+a_1\log P}$ obtained from a bootstrap re-sampling experiment for 
the sub-samples of extremely-blue (XB, continuous line) and moderately-blue 
(MB, broken line) systems. These distributions have been used to establish the 
uncertainty estimates given in Table~1.} 
\end{figure}
Histograms of the individual values of the zero-point and slope coefficients 
from the bootstrap experiment are shown in Fig.~5. Note that the mean color 
indices for the XB and MB groups are ${\overline{V-R_C}=-0.142}$ and ${-
0.047}$, while the difference of the zero points is: ${a_0({\rm MB})-a_0({\rm 
XM})=0.84}$. This would imply a very strong dependence on the color index with 
a steep slope of about 8.8, but part of this is almost certainly contributed 
by the relatively larger reddening for the MB group. Much more interesting are 
the period-dependence slope coefficients $a_1$. The determination of $a_1$ for 
the XB group is surprisingly stable, given the large scatter of the data 
points and the fact that the value of $a_1$ is driven mostly by the outlying 
points, ${a_1({\rm XB})=-2.74^{+0.58}_{-0.68}}$; the range given here is 
equivalent to the {\it rms} error of 1-sigma. The determination for the MB 
group is poorer, ${a_1({\rm MB})=-2.07^{+0.83}_{-1.00}}$, probably because the 
presence of the faint cut-off in the data and certainly stronger influence of 
the scatter in the reddening values for this group. When compared with the 
calibrations for the W~UMa-type binaries using the ${B-V}$ and ${V-I_C}$ 
indices (\cf Eqs.~(2) and (5a) in Rucinski and Duerbeck 1997), the slope 
${a_1({\rm XB})=-2.74}$ is rather shallow (the previous calibrations implied 
the slope of about ${-4.4}$) which may have resulted from relatively large 
photometric errors of the MACHO data. 

\Section{Conclusions for the Future}
The main result of this pilot study is that the sample of EB3 binaries 
discovered and classified by the MACHO project certainly contains hot, blue, 
massive binaries of the contact type. Although it may have been preferable to 
make the selection of the systems on the basis of the Fourier coefficients -- 
rather than to rely on the MACHO classifications -- this was not necessary 
because the systems reveal typical properties of contact-binary stars. The 
contact nature of these binaries is most strongly confirmed by the existence 
of a period--luminosity relation which is best visible in the XB sub-sample 
consisting of the bluest and least reddened systems. The new result is that 
the massive, young, blue systems in LMC are apparently not constrained by the 
previously-established short-period blue envelope in the period--color diagram 
and appear with blue-color/short-period combinations not observed in the 
previous surveys of the open clusters in the Galaxy and in the Galactic Disk 
field. These systems appear also within the orbital period interval of 1.3--1.5 
to 2--3 days, a range for which the frequency of the galactic contact systems 
is known to be un-measurably low. 

An important conclusion related to the future attempts of determining an 
absolute-magnitude calibration for early-type contact systems is the 
availability of accurate color indices. Photometry should be available in at 
least three bandpasses for reddening determinations, and must be accurate 
enough (through the use of the standard filter bands and a requirement that  
mean standard errors be ${<0.01}$~mag) to compensate for the decreased 
sensitivity of optical color indices to effective temperature. For comparison 
and consistency checks with the results for solar-type contact binaries of the 
W~UMa-type, it would be preferable to utilize ${B-V}$ and ${V-I_C}$ color 
indices. The new database for SMC coming from the OGLE-II project (Udalski 
\etal 1998) apparently fulfills most of the above desiderata. 

\Acknow{Thanks are due to Bohdan Paczynski for pointing the existence of the 
MACHO data for eclipsing binaries in LMC and for very useful comments and 
suggestions on the first version of the paper.}

\end{document}